\documentclass[12pt]{article}    
\usepackage{amsfonts,graphicx,epsfig} 

\begin{document}


\newcommand{\0}{\mbox{\boldmath $0$}}         
\newcommand{\ba}{\mbox{\boldmath $a$}}        
\newcommand{\balpha}{\mbox{\boldmath $\alpha$}} 
\newcommand{\bbeta}{\mbox{\boldmath $\beta$}} 
\newcommand{\bd}{\mbox{\boldmath $d$}}        
\newcommand{\beps}{\mbox{\boldmath $\epsilon$}} 
\newcommand{\binom}[2]  {{{#1}\choose{#2}}}   
\newcommand{\bj}{\mbox{\boldmath $j$}}        
\newcommand{\bJ}{\mbox{\boldmath $J$}}        
\newcommand{\blam}{\mbox{\boldmath $\lambda$}}
\newcommand{\bpi}{\mbox{\boldmath $\pi$}}     
\newcommand{\bp}{\mbox{\boldmath $p$}}        
\newcommand{\bsig}{\mbox{\boldmath $\sigma$}} 
\newcommand{\bth}{\mbox{\boldmath$\theta$}}   
\newcommand{\dt}{\widetilde{d}}               
\newcommand{\E}{{\rm E}}                      
\newcommand{\m}{\mbox{\boldmath $m$}}         
\newcommand{\M}{\mbox{\boldmath $M$}}         
\newcommand{\mi}{{$-$}}                       
\newcommand{\n}{\mbox{\boldmath $n$}}         
\newcommand{\ndot}{n{\rm\bf .}}               
\newcommand{\Nh}{\widehat{N}}                 
\newcommand{\p}{\phantom{5}}                  
\newcommand{\pih}{\widehat{\pi}}              
\newcommand{\pmi}{\phantom{$-$}}              
\newcommand{\ps}{\phantom{$*$}}               
\newcommand{\pc}{{\scriptsize$\bullet$}}      
\newcommand{\q}{\phantom{55}}                 
\newcommand{\qs}{\hspace*{.1in}}              
\newcommand{\re}{{\rm e}}                     
\newcommand{\U}{\mbox{\boldmath $u$}}         
\newcommand{\X}{\mbox{\boldmath $X$}}         
\newcommand{\x}{\mbox{\boldmath $x$}}         
\newcommand{\Y}{\mbox{\boldmath $Y$}}         
\newcommand{\y}{\mbox{\boldmath $y$}}         
\newcommand{\z}{\mbox{\boldmath $z$}}         
\newcommand{\bgam}{\mbox{\boldmath $\gamma$}}

\hyphenation{general-iza-tions}
\hyphenation{distri-bution}


\thispagestyle{empty}
\begin{center}
\vspace*{\fill}

{\Large\bf   Distributions Associated With} \\[1ex]
{\Large\bf   Simultaneous Multiple Hypothesis Testing} \\[1ex] 

\vspace*{1.5in}
\begin{tabular}{l}
 
{\bf Chang Yu}  \\[.25in]
Department of Biostatistics  \\
Vanderbilt University Medical Center       \\
Nashville, TN 37232  \\[.2in]

{\bf Daniel Zelterman}${}^*$  \\[.25in]
Department of Biostatistics  \\
Yale University       \\
New Haven, CT  06520  \\[.2in]

\today
\end{tabular}
\end{center}
\vspace*{\fill}

\noindent${}^*$Email address for corresponding author: {\tt daniel.zelterman@yale.edu}. 
The authors thank J.~Jin and T.~Cai for providing the data analyzed in Section~\ref{breast.cancer.section}.
This work was supported in part by Vanderbilt CTSA grant 1ULTR002243 from NIH/NCATS, R01 CA149633 from NIH/NCI, R01 FD004778 from FDA, R21 HL129020, P01 HL108800 from NIH/NHLBI (CY) and grants P50-CA196530, P30-CA16359, R01-CA177719, R01-ES005775, R41-A120546, U48-DP005023, and R01-CA168733 awarded by NIH (DZ).

\newpage
\thispagestyle{empty}

\section*{Abstract}

We develop the distribution of the number of hypotheses found to be statistically significant using the rule from Benjamini and 
Hochberg~(1995)\label{BH.1995.1} for controlling the false discovery rate (FDR).
This distribution has both a small sample form and an asymptotic expression for testing many independent hypotheses simultaneously.
We propose a parametric distribution $\,\Psi_I(\cdot)\,$ to approximate the marginal distribution of p-values under a non-uniform alternative hypothesis. 
This distribution is useful when there are many different alternative hypotheses and these are not individually well understood.
We fit $\,\Psi_I\,$ to data from three cancer studies and use it to illustrate the distribution of the number of notable hypotheses observed in these examples. 
We model dependence of sampled p-values using a copula model and a latent variable approach.
These methods can be combined to illustrate a power analysis in planning a large study on the basis of a smaller pilot study.
We show the number of statistically significant p-values behaves approximately as a mixture of a normal and the Borel-Tanner distribution.

\bigskip

\noindent{\bf Keywords:} Benjamini-Hochberg criteria; 
Borel-Tanner distribution; false discovery rate; order statistics; p-value

\thispagestyle{empty}
\setcounter{page}{1}
\section{Introduction}
\typeout{}
\typeout{Section \thesection: Introduction}
\typeout{}

Much work in informatics is concerned with identifying and classifying statistically significant biological markers.
In this work we develop methods for describing the distribution of the numbers of such events.

Informatics methods often summarize experiments resulting in a large number of p-values, usually through multiple comparisons of gene expression data. 
Typically, the number of tests $\,n$, is much greater than the number of subjects.
There are popular rules for identifying statistically significant p-values while maintaining a false discovery rate (FDR) below a pre-specified level $\,\alpha\; (0<\alpha<1)$.
Benjamini~(2010)\label{Benjamini.2010} provides a review of recent advances.

A commonly cited rule to control the experiment-wise FDR is the  Bonferroni correction. 
Given a sample of ordered p-values $\,p_1\leq p_2 \leq \cdots \leq p_n$, the Bonferroni rule finds the smallest value of $\,B=0,1,\ldots, n-1\,$ for which
\begin{equation}                              
          p_{B+1} > \alpha/n \; .             \label{B.def.eq}
\end{equation}

The rule developed by Benjamini and Hochberg (1995)\label{BH.1995.2} to maintain a FDR $\leq\alpha\,$ finds the smallest value of $\,k\,$ (denoted by BH) such that
\begin{equation}                              
   p_1\leq\alpha/n;\;\; p_2\leq2\alpha/n;\; \cdots\; p_k\leq k\alpha/n;
{\rm \ \ and \ \ } p_{k+1} \; > \; (k+1)\alpha/n\; .  \label{BH.def.eq}
\end{equation}

We describe the probability of  BH=$\,k\,$ under independent null hypotheses where each p-value has a marginal uniform distribution as well as an alternative approximating  distribution with density function $\,\psi_I(p)\,$ expressible as a polynomial in $\,\log p$ of order $\,I$.
We will also examine this distribution accounting for dependence among the p-values in Section~\ref{dependence.section}.

There has been little research on parametric distributions for the p-values generated from data under a mixture of the null and multiple  alternative hypotheses.  
The mixed p-values are mainly modeled using non-parametric methods (Genovese and Wasserman, 2004;\label{Genovese.2004.1} Broberg, 2005;\label{Broberg.2005.1} Langaas, Lindqvist, and Ferkingstad, 2005;\label{Langaas.2005.1} Tang, Ghosal, and Roy, 2007)\label{Tang.2007.1} or alternatively, the p-values are converted into normal quantiles and modeled thereafter (Efron, Tibshirani {\em et al.}, 2001;\label{Efron.2001} 
Efron, 2004;\label{Efron.2004.1} Jin and Cai, 2007).\label{Jin.2007.1}   
Another common approach is to approximate the distribution of sampled p-values using a mixture of beta distributions (Pounds and Morris 2003;\label{Pounds.2003.1} Tang, Ghosal, and Roy 2007).

One aim of this work is to propose a parametric distribution for 
p-values independently of the statistical tests used to generate them.  
Another use for the proposed distribution $\,\Psi_I\,$ is to estimate the proportion $\,\pi_0\,(0\leq\pi_0\leq 1)\,$ of p-values sampled from the null hypothesis.
If all of the empirical p-values are generated under the null hypothesis $\,(\pi_0 = 1)\,$ then these are well-known to follow a uniform distribution.   
We are also interested in a setting where a fraction $\,1 - \pi_0\,$ of the tests are performed under a variety of alternative hypotheses. 
Benjamini and Hochberg~(2000)\label{BH.2000} recommend we perform tests with significance level $\,\alpha/\pi_0\,$ and still maintain a FDR below $\,\alpha$.
Langaas, Lindqvist, and Ferkingstad (2005)\label{Langaas.2005.2} and Tang, Ghosal, and Roy (2007) suggest the estimated density of p-values at $\,p=1\,$ be used to estimate the fraction $\,\pi_0\,$ of p-values sampled under the null hypothesis. 
We found $\,\psi_I(1\mid\widehat{\bth})\,$ as a useful estimator of $\,\pi_0\,$ in the examples of Section~\ref{examples.section}, where
 $\,\widehat{\bth}\,$ denote maximum likelihood estimates.

The p-values are usually not independent.   
In microarray studies, for example, a small number of clusters of 
p-values in the same biological pathway will have high mutual correlations.  
Methods for modeling such dependencies are developed by Sun and Cai~(2009), Friguet, Kloareg, and Causeur~(2009),\label{Friguet.2009} and Wu~(2008)\label{Wu.2008} for examples.

In Section~\ref{basic.results.section} we describe the probability distribution of BH in~(\ref{BH.def.eq}) when the $\,p_i\,$ are independently sampled from a uniform distribution under the null hypothesis and from an unspecified distribution $\,\Psi\,$ under the alternative.
Section~\ref{Psi.section} provides elementary properties of the proposed  distribution $\,\Psi_I$.   
The inclusion of $\,\Psi_I\,$ allows us to easily approximate the behavior of p-values under the alternative hypothesis, facilitating modeling of the distribution of the number of identified p-values under the alternative hypothesis.
The parameters $\,\bth\,$ of $\,\Psi_I\,$ depend on the specific application and are estimated for three examples in Section~\ref{examples.section}.
In Section~\ref{dependence.section} we describe the distribution of BH modeling dependence of p-values using two approaches: sampling from a copula model; and conditioning on a latent variable. 
We combine these methods in Section~\ref{power.section} to illustrate approximate power in planning a proposed study involving multiple hypothesis testing settings.
In Appendixes~A and~B we provide details of the distribution of BH as a mixture of Borel-Tanner and normal approximating distributions.
Appendix~C examines the parameter space for the $\,\Psi_I\,$ distribution.

\newpage\section{Simultaneous Multiple Testing}
\typeout{}
\typeout{Section \thesection: Simultaeous Multiple Testing}
\typeout{}
\label{basic.results.section}

Let $\,p_1\leq p_2\leq\cdots\leq p_n\,$ denote the ordered $\,p-$values sampled from an exchangeable parent population and let $\,f_k\,$ denote the joint density function of any $\,k\,$ of these.
Then the probability of event~(\ref{BH.def.eq}) is 
\begin{eqnarray}
 \Pr[\,{\rm BH}=k\,] &=& \frac{n!}{(n - k)!}      
                 \label{BH.integral.eq} \\
      &&  \hspace*{-.75in} \times\int_{p_1 = 0}^{\alpha / n}
         \int_{p_2 = p_1} ^ {2\alpha / n}
                 \;\cdots\;\nonumber
         \int_{p_k = p_{k - 1}} ^ {k \alpha / n}
        \int_{p_{k + 1} = (k + 1) \alpha / n} ^ 1 
           f_{k + 1}(p_1, p_2, \ldots , p_{k + 1})
           \,{\rm d}p_{k+1}\ldots\, {\rm d}p_2 \,{\rm d}p_1 .
\end{eqnarray}

The Bonferroni rule replaces the upper limits with the same value $\,\alpha/n\,$ on all but the innermost of these integrals.  
The range of the innermost integral for the Bonferroni rule is from $\,\alpha/n\,$ to 1.

In Section~\ref{Psi.section}, we demonstrate the assumption all $\,p_i\,$ have the same distribution may not be critical.
For the remainder of the present section, let us make the additional assumption of independence of the parent population of p-values.
In Section~\ref{dependence.section} we will return to this assumption of independence and describe dependence of all p-values under two sampling models.
If the p-values are independent then we can use well-known results about order statistics.

Let $\,\Psi(\cdot)\,$ denote the marginal distribution function of the $\,p_i\,$ with corresponding density function $\,\psi(\cdot)$.
Then
$$
     f_k (p_1,\ldots,p_k) = \psi(p_1)\cdots\psi(p_k)\; .
$$

If the $\,p_i\,$ are sampled from independent null hypotheses then $\,\Psi\,$ is uniform.
In Section~\ref{Psi.section} we propose an approximation for $\,\Psi\,$ under a non-uniform alternative hypotheses and use this to develop~(\ref{BH.integral.eq}).

If we follow the Bonferroni rule~(\ref{B.def.eq}) then the distribution of the number of statistically significant p-values at FDR$\,\leq\alpha\,$ follows a binomial distribution with index $\,n\,$ and probability parameter equal to $\,\Psi(\alpha / n)$.

Let BH denote the number of statistically significant p-values at FDR$\,\leq\alpha\,$ using the Benjamini-Hochberg criteria~(\ref{BH.def.eq}).
Then~(\ref{BH.integral.eq}) gives
\begin{eqnarray*}
   \Pr[\,{\rm BH}=0\,] &=&  
    \Pr[\, {\rm all \ } p_i > \alpha / n  \,] \nonumber \\
    &=& \{ 1 - \Psi(\alpha / n)\} ^ n \; ,\\
\Pr[\,{\rm BH}=1\,] &=&    \nonumber
    \Pr[\, p_1 \leq \alpha / n ;\; 
        {\rm all \ other \ } p_i > 2\alpha / n \,] \\
    &=& n\Psi(\alpha / n)\{1 - \Psi(2 \alpha / n)\} ^ {n - 1} \; , 
\end{eqnarray*}
\noindent and integrating terms in~(\ref{BH.integral.eq}) gives  
\begin{eqnarray*}
\Pr[\,{\rm BH}=2\,] &=&
    \Pr[\, p_1 \leq \alpha / n;\; p_1 \leq p_2 \leq 2 \alpha / n;\;
       {\rm all \ other \ } p_i > 3 \alpha / n\,] \nonumber \\
      &=& n (n - 1) \Psi(\alpha / n)
        \{\Psi(2 \alpha / n) - \Psi(\alpha / n) / 2\} 
                \{1 - \Psi(3 \alpha / n)\} ^ {n - 2}\, .
\end{eqnarray*}

The general form, for $\,k = 0, 1, 2, \ldots , n\,$ is
\begin{equation}              \label{BH.eq}   
\Pr[\,{\rm BH}=k\,] = \frac{n!} {(n - k)!}  \, U_k \,
         \{ 1 - \Psi((k+1)\alpha / n)\}^{n-k}  
\end{equation}
where $\,U_0 =1\,$ and
\begin{equation}                \label{Uk.eq}
  U_k = \int_{p_1=0}^{\alpha/n} \int_{p_2= p_1}^{2\alpha/n}
       \;\cdots\;\int_{p_k = p_{k-1}}^{k\alpha/n}
        \psi(p_1)\,\psi(p_2)\,\cdots\,\psi(p_k)\,  
        {\rm d}p_k\,{\rm d}p_{k-1}\ldots {\rm d}p_1
\end{equation}
for $\,k=1,\ldots ,n$.

In Appendix~A we show $\,U_k\,$ can be evaluated according to the recursive form
\begin{equation}                   \label{Uk.recursive.eq}
    U_k = \sum_{i=1}^k \; (-1)^{i+1} \;       
                \Psi^i\{(k-i+1)\alpha/n\} \,U_{k-i}\,/\,i!
\end{equation}
facilitating its numerical evaluation for the examples in Section~\ref{examples.section}.

Let us next examine the special case where all p-values are independently sampled under the null hypothesis.
When the marginal distribution of the $\,p_i\,$ are independent and uniformly distributed (that is, $\,\Psi(p) = p\,$) then~(\ref{BH.eq}) is expressible as polynomials in $\,\alpha$.

Specifically,
\begin{eqnarray*}
   \Pr[\, {\rm BH} = 0\,] &=& (1 - \alpha / n) ^ n \\
\Pr[\, {\rm BH} = 1\,] &=& \alpha \,(1- 2 \alpha / n) ^ {n-1} \\
\Pr[\, {\rm BH} = 2\,] &=& 3/2\;\{(n-1)/n\} \,\alpha^2\, 
                            (1-3\alpha/n)^ {n - 2} 
\end{eqnarray*}
and in general,
\begin{equation}                       
\Pr[\, {\rm BH} = k\,] =
  {{n}\choose{k}} \; (k+1)^{k-1} \,(\alpha/n)^k \label{exact.eq}
     \,\{1 - (k+1)\alpha/n\}^{n-k}\; . 
\end{equation}
Details of the derivation of (\ref{exact.eq}) appear in Appendix~A.

Useful results can be obtained if we also assume the number of independent hypotheses $\,n\,$ is large.
In this setting, the number of identified p-values at FDR$\,\leq\alpha\,$ for the Bonferroni criteria~(\ref{B.def.eq}) will follow a Poisson distribution with mean $\,\alpha$.

The limiting probabilities of the BH distribution at~(\ref{exact.eq}) for large $\,n\,$ are as follows:
\begin{eqnarray*}
  \Pr[\,{\rm BH } = 0\,] & = & e^{-\alpha}  \\
  \Pr[\,{\rm BH } = 1\,] & = & \alpha \,e^{-2\alpha}  \\
\hspace*{-2.75in}{\rm and} \hspace{2.75in}&&\\
  \Pr[\,{\rm BH } = 2\,] & = & 3/2 \; \alpha^2\,e^{-3\alpha} \; .
\end{eqnarray*}

The general form for the limiting probabilities in~(\ref{exact.eq}) is 
\begin{equation}                        
  \Pr[\,{\rm BH} = k\,] = \{(k+1)^{k-1}/\,k!\,\}
           \;\alpha^k \;e^{-(k+1)\alpha} \label{Borel.eq}
\end{equation}
for $\,k=0,1,\ldots\,$ but much smaller than $\,n$.

The probabilities in~(\ref{Borel.eq}) sum to unity using the relation in Jolley (1961, eqn.~(130), p.~24).\label{Jolley.1961}
The moments of this distribution can be obtained by successively differentiating both sides of the relation $\,\sum\Pr[{\rm BH}=k]=1\,$ with respect to $\,\alpha$.
Specifically, the mean of~(\ref{Borel.eq}) is $\,\alpha/(1-\alpha)\,$ and the variance is $\,\alpha / (1 - \alpha)^3$.
If all p-values are sampled from the uniform null hypothesis, then (\ref{Borel.eq}) shows
$$
      \Pr[\,{\rm BH} \geq 1\,] = 1-e^{-\alpha} < \alpha
$$ 
so the FDR$\leq \alpha\,$ is maintained for the BH procedure.
The distribution of BH+1 in~(\ref{Borel.eq}) is known as the {\it Borel-Tanner distribution\/} with applications in queueing theory (Tanner, 1961).\label{Tanner.1961}

\section{Distributions for $\,P-$Values}
\typeout{}
\typeout{Section \thesection: P-values under the alternative}
\typeout{}
\label{Psi.section}

We need a marginal distribution for p-values, independent of the choice of test statistic.
We continue to assume the p-values are mutually independent and have the same marginal distrbutions.
We must have $\,\Psi\,$ concave (Genovese and Wasserman (2004),\label{Genovese.2004.2} Sun and Cai (2009))\label{Sun.2009} otherwise the underlying test will have power smaller than its significance level for some $\,\alpha$.
Similarly, the corresponding density function $\,\psi\,$ must be monotone decreasing.
Beta distributions and mixtures of betas have been proposed for this purpose by Pounds and Morris~(2003),\label{Pounds.2003.2} Broberg~(2005),\label{Broberg.2005.2} and Tang, Ghosal, and Roy~(2007),\label{Tang.2007.2} among others.
Other parametric models have been proposed by Kozoil and Tuckwell~(1999),\label{Kozoil.1999} Genovese and Wasserman~(2004),\label{Genovese.2004.3} and Yu and Zelterman~(2017).\label{Yu.2017}
We next propose a different flexible distribution for modeling p-values under alternative hypotheses.

Consider a distribution with a density function expressible as a polynomial in $\,\log p\,$ up to degree $\,I=0,1,2,\ldots$.
The uniform (0--1) distribution is obtained for $\,I=0$.
The marginal density function we propose for p-values is
\begin{equation}                       
    \psi_I (p\mid\bth) =                       \label{psi.eq}
            \sum_{i=0}^I\; \theta_i\,(-\log p)^i 
\end{equation}
for real-valued parameters $\,\bth = \{\theta_1, \ldots , \theta_I\}\,$ with $\,I\geq1\,$ where
\begin{equation}                        
     \theta_0 = 1-\sum_{i=1}^I \; i!\theta_i   \label{theta0.eq}
\end{equation}
so the densities $\,\psi_I(p)\,$ integrate to one.

The corresponding cumulative distribution function is
\begin{equation}                        
     \Psi_I(p\mid \bbeta) = p\;\sum_{i=0}^I \; 
          \beta_i\, (-\log p)^i               \label{Psi.eq}        
\end{equation}
where $\,\beta_0=1$.

The relationships between these parameters is
$$
             \beta_j = \sum_{i=j}^I \; \theta_i \,i!/j!
$$
for $\,j=1,2, \ldots, I\,$ and 
$\, \theta_i = \beta_i - (i+1)\beta_{i+1}\, $
for $\,i=1,2, \ldots, I-1$.
Throughout, we will interchangeably refer to either the $\,\bth\,$ or $\,\bbeta\,$ parameterizations.

The moments of distribution $\,\psi_I(p\mid\bth)\,$ are
\begin{equation}                              
    \E(p^{\,j}\mid\bth) =     \label{Psi.moment.eq}
       \sum_{i=0}^I\; i!\,\theta_i  \,/ \, (j+1)^{i+1} 
\end{equation}
for $\,j=1,2,\ldots\,$.

We must have $\,\theta_I>0\,$ in order to have $\,\psi_I(p)>0\,$ for values of $\,p\,$ close to zero.
Values of $\,\theta_0\,$ are restricted at~(\ref{theta0.eq}) in order for $\,\psi_I(p)\,$ to integrate to unity.
Since $\,\psi_I(1\mid\bth) = \theta_0\,$ we must also require $\,\theta_0\geq 0$.
Requiring $\,\psi(p)\,$ to be decreasing at $\,p=1\,$ gives $\,\theta_1\geq 0$.

These restrictions alone on $\,\theta_0\;\theta_1,\,$ and $\,\theta_I\,$ are not sufficient to guarantee $\,\psi_I(p\mid\bth)\,$ is monotone decreasing or positive valued for all values of $\,0\leq p\leq 1$. 
The necessary conditions for achieving these properties are difficult to describe in general but it is sufficient that all $\,\theta_i\geq0$.  
Specific cases are examined in Appendix~C for values of $\,I\,$ up to $\,I=4$.
Models for larger values of $\,I\,$ could be fitted by maximizing the penalized likelihood, such that $\,\psi_I(p\mid\bth)\,$ is positive valued and monotone decreasing at the observed, sorted p-values.

In practice, the choice of $\,I\,$ is found by fitting a sequence of models.
Successive values of $\,I\,$ represent nested models so twice the differences of log-likelihoods will behave as $\,\chi^2\,$ (1~df) when the underlying additional parameter value is zero.
In the examples of Section~\ref{examples.section}, we found $\,I=3\,$ or 4 produced an adequate fit.

The $\,\psi_I\,$ density function is specially suited for modeling the marginal distribution of a variety of non-uniform distributions for p-values.
If each $\,p_i\, (i=1,\ldots ,n\,$) is sampled from a different distribution with density function $\,\psi_I(p\mid\bth_i)\,$ then the marginal density of all $\,p_i\,$ satisfies 
\begin{equation}                          
      n^{-1} \sum^n_i \psi_I(p \mid \bth_i)     \label{mixture.eq}
        = \psi_I(p \mid \overline {\bth}),
\end{equation}
where $\,\overline{\bth}\,$ is the arithmetic average of all $\,\bth_i$.
A similar result holds if the values of $\,I\,$ vary across distributions of $\,p_i$.

This mixing of distributions includes the uniform as a special case.
Specifically, suppose $\,100\pi_0-$percent of the p-values are sampled from a uniform (0, 1) distribution $(0 \leq\pi_0 \leq 1$) and the remaining $\,100(1-\pi_0)-$percent are sampled from 
$\,\psi_I(p\mid\bth)$.
Then the marginal distribution has density function
\begin{equation}                          
    \pi_0 + (1-\pi_0)\,\psi_I(p\mid\bth)
        \; =\; \psi_I(p\mid(1-\pi_0)\bth)           \label{Umix.eq}
\end{equation}
demonstrating $\,\pi_0\,$ is not identifiable in this model.

Equations~(\ref{mixture.eq}) and~(\ref{Umix.eq}) illustrate the utility of $\,\psi_I\,$ in modeling p-values sampled from a mixture of the null hypothesis and many different alternative hypotheses and yet retaining the same parametric distribution.
Donoho and Jin~(2004)\label{Donoho.2004} also describe the value of such a mixture of heterogeneous alternative hypotheses in multiple testing settings.

Following Langaas, Lindqvist, and Ferkingstad (2005)\label{Langaas.2005.3} and Tang, Ghosal, and Roy (2007),\label{Tang.2007.3} we use the estimated density at $\,p=1\,$ to estimate $\,\pi_0$, the proportion of p-values sampled from the null hypothesis.
The estimated values of $\,\psi_I(1\mid\widehat{\bth}) = \widehat\theta_0\,$ are given in Table~\ref{fitted.table} for all fitted models and examples of the following section.

\section{Application}
\label{examples.section}
\typeout{}\typeout{Section \thesection: Two Examples}
\typeout{}

For each of three examples we fitted the density function $\,\psi_I\,$ described in Section~\ref{Psi.section} and then used these to examine the distribution of BH at~(\ref{BH.eq}).
The fitted parameter values $\,\widehat{\bth}\,$ for each of these examples appear in Table~\ref{fitted.table} for successive values 
of $\,I$.
We maximized the likelihoods using standard optimization routine {\bf nlm} in {\bf R}.
This routine also provides estimates of the Hessian used to estimate standard errors of parameter estimates in Table~\ref{fitted.table}.

In these examples, the evaluation of $\,U_k\,$ in~(\ref{Uk.recursive.eq}) involves adding and subtracting many nearly equal values resulting in numerical instability.
We computed~(\ref{BH.eq}) using multiple precision arithmetic with the {\bf Rmpfr} package in {\bf R}.
Table~\ref{fitted.table} also displays the fitted parameters for a third example, introduced in Section~\ref{power.section}, to illustrate estimation of power for multiple hypothesis testing problems.

\begin{table}[p]
\begin{center}
\caption{Maximum likelihood estimated parameter values of $\,\psi_I\,$ for three examples.  \label{fitted.table}}
\begin{tabular}{cccc ccc}   \\ \hline \\[-2.75ex]
\multicolumn{4}{c}{Model parameters} 
       & \mi Log- & $2\times$ &  $\,\widehat\theta_0\,$ to  \\
       \cline{1-4} \\[-2.5ex]
$I$ & Symbol & Estimate & Std Err  & Likelihood 
        & Difference  
        &  estimate $\,\pi_0$
        \\  \hline
 \multicolumn{7}{c}{Section~\ref{breast.cancer.section}: BRCA in breast cancer, $\,n=3226$} \\ \hline
 1 & $\theta_1$ & 0.531 & 0.018 & 482.04 & --- & 0.469 \\[1ex]
 2 & $\theta_1$ & 0.0\q & 0.049 & 569.04 & 174.0 & 0.649 \\
   & $\theta_2$ & 0.177 & 0.015 & \\[1ex]
 3 & $\theta_1$ & 0.158 & 0.084  & 573.27 & \q8.47 & 0.623\\
   & $\theta_2$ & \p0.0492 & \p0.0506 \\
   & $\theta_3$ & \p0.0201 & \p0.0075 \\[1ex]
 4 & \multicolumn{4}{l}{\quad Same as $\,I=3$}\\ \hline
 \multicolumn{7}{c}{Section~\ref{lung.section}: Smoking and squamous cell lung cancer, $\,n=20,068$} 
     \\ \hline
 1 & $\theta_1$ & 0.448 & 0.007 &  2147.48 & --- & 0.552\\[2ex]
 2 & $\theta_1$ & 0.0\q & 0.020 & 2579.47 & 863.98 & 0.684\\
   & $\theta_2$ & 0.158 & 0.006\\ [1ex] 
 3 & $\theta_1$ & 0.174 & 0.034 & 2641.32 & 123.70 & 0.684 \\
   & $\theta_2$ & \p0.0008 & 0.020 \\
   & $\theta_3$ & \p0.0233 & \p0.0028 \\[1ex]     
 4 & $\theta_1$ & 0.100 & \p0.0497 & 2643.49 & \q4.33 & 0.698 \\
   & $\theta_2$ & \p0.0761 & \p0.0423\\
   & $\theta_3$ & \p\q0.000493 & \p0.0119\\
   & $\theta_4$ & \q0.00195 & \p0.0010 \\[1ex]
 5 & \multicolumn{5}{l}{\quad Same as $\,I=4$} \\ \hline
  \multicolumn{7}{c}{Section~\ref{power.section}: Survival in lung cancer: $\,n=48,803$ markers; $N=78$ patients} \\ \hline 
 1 & $\theta_1$ & 0.1366\p & 0.0048 & 461.09 & --- & 0.863 \\[1ex]
 2 & $\theta_1$ & 0.00863 & 0.0115 & 541.72 & 161.26  & 0.921\\
   & $\theta_2$ & 0.03507 & 0.0030 \\[1ex]
 3 & $\theta_1$ & 0.0524\p & 0.020\p & 545.44 & \q7.45  & 0.908\\
   & $\theta_2$ & 0.00983 & 0.010\p \\
   & $\theta_3$ & 0.00327 & 0.0013 \\[1ex]
 4 &   \multicolumn{5}{l}{\quad Same as $\,I=3$}    \\ \hline
\end{tabular}                                             
\end{center}
\end{table}

\subsection{Breast Cancer}
\label{breast.cancer.section}

This microarray dataset was originally described by Hedenfalk, Duggan {\em et al}~(2001)\label{Hedenfalk.2001} and also analyzed by Storey and Tibshirani~(2003).\label{Storey.2003}
These data summarize marker expressions of 3226 genes in seven women with the BRCA1 mutation and in eight women with the BRCA2 mutation.  
The objective was to determine differentially-expressed genes between these two groups. 
Earlier analyses used a two-sample t-test to compare the two groups for each gene, giving rise to $\,n=3226\,$ p-values. 
Efron~(2004)\label{Efron.2004.2} and Jin and Cai~(2007)\label{Jin.2007.2} model the z-scores corresponding to the p-values.

\typeout{}\typeout{Observed and fitted breast cancer p-values}\typeout{}
\begin{figure}[t]
\vspace*{-.5in}
\begin{center}
\includegraphics[scale=.7]{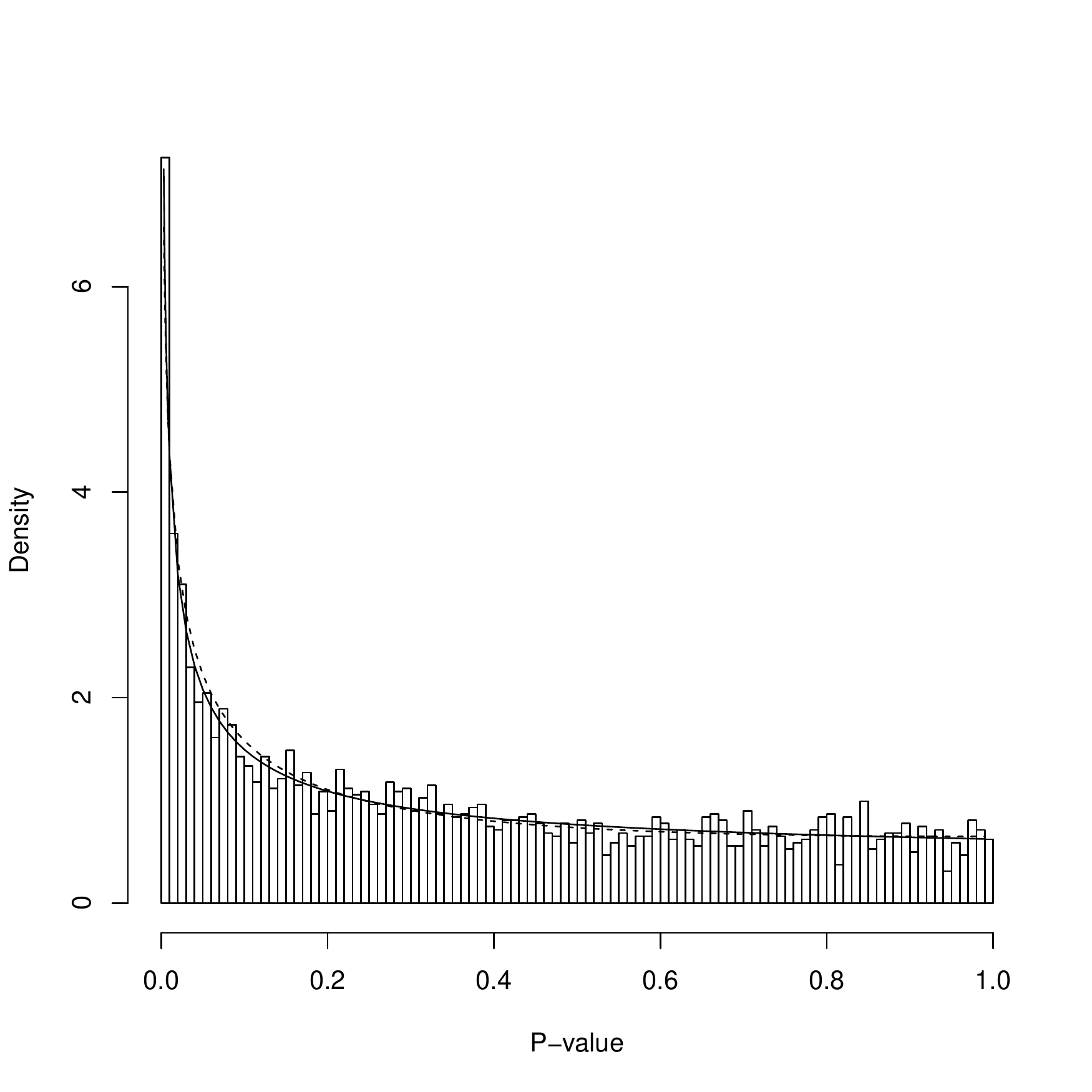}
\end{center}
\vspace*{-.35in}
\caption{Observed and fitted density function $\,\psi_I\,$ for the 3226 p-values from the breast cancer data with $\,I=2\,$ (dashed line) and $\,I=3\,$ (solid line). Maximum likelihood parameter estimates are given in Table~\ref{fitted.table}. \label{breast.cancer.figure}}
\vspace*{.35in}
\end{figure}
\typeout{}\typeout{Breast cancer fitted BH model}\typeout{}
\begin{figure}[h]
\vspace*{-.45in}
\begin{center}
\includegraphics[scale=.65]{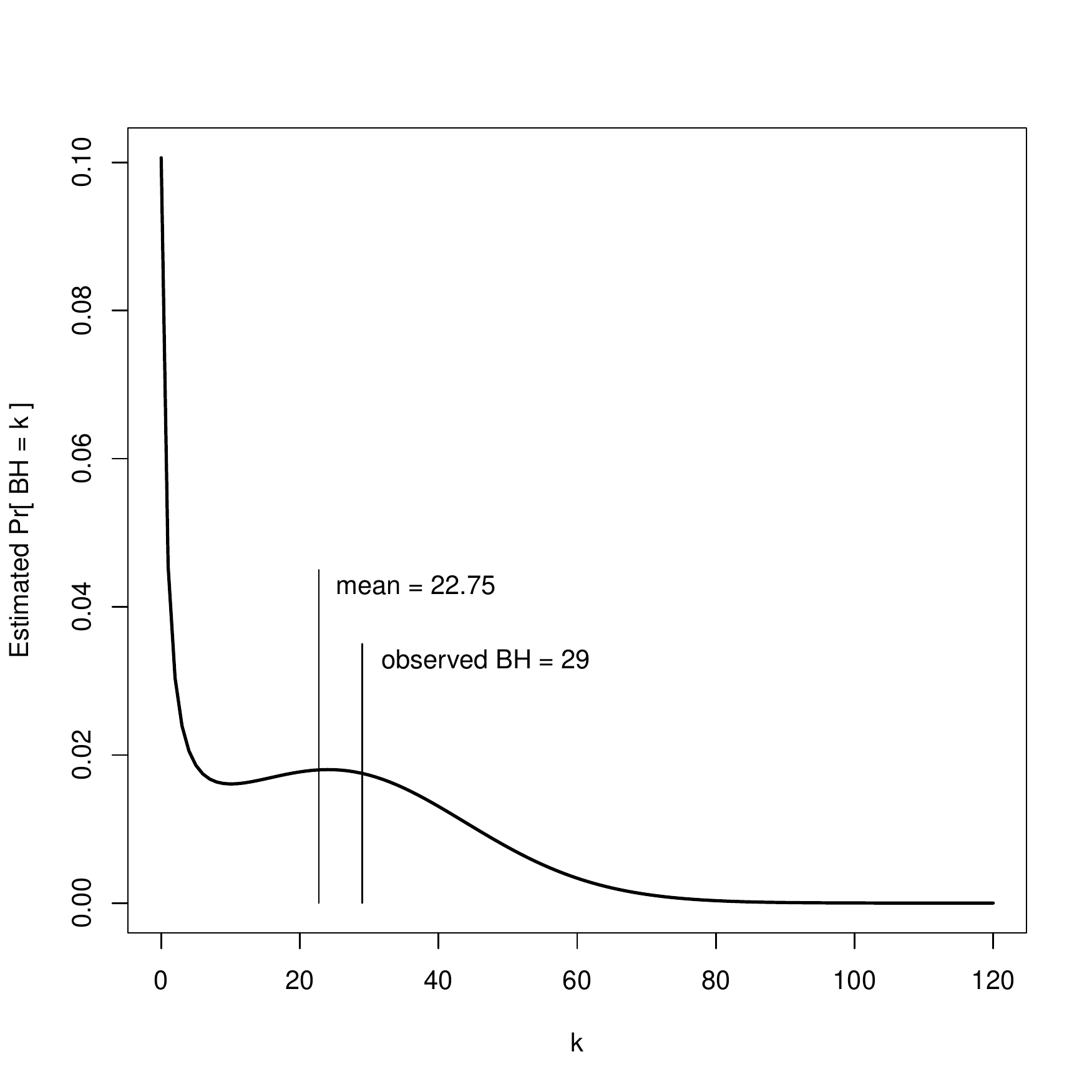}
\end{center}
\vspace*{-.35in}
\caption{Fitted BH distribution for breast cancer p-values with $\,I=3$.  The expected value and observed BH = 29 p-values at FDR$\,=.05\,$ are indicated.  \label{BCBH.fig}}
\vspace*{.35in}
\end{figure}

The maximum likelihood fitted values $\,\widehat{\bth}\,$ for $\,\psi_I\,$ are given in Table~\ref{fitted.table}.  
The model for $\,I=2\,$ represents a big improvement over the model with $\,I=1\,$ parameters.
The model with $\,I=3\,$ parameters has a modest improvement over the model with $\,I=2\,$ and $\,I=4\,$ demonstrates negligible change in the likelihood over $\,I=3$.
Fitted densities $\,\psi_I\,$ for $\,I=2\,$ and 3 are plotted in Fig.~\ref{breast.cancer.figure} along with the observed data. 
There is small difference between the fitted models in this figure and both exhibit a good fit to these data.
Our estimate of $\,\pi_0\,$ given by $\,\widehat\theta_0\,$ is .65 for $\,I=2\,$ and .62 for $\,I=3$.

There are BH=29 statistically significant markers at FDR$\,=\,.05$ using the adjustment for multiplicity at~(\ref{BH.def.eq}).
The fitted BH distribution~(\ref{BH.eq}) is displayed in Fig.~\ref{BCBH.fig} using the fitted parameters $\,\widehat{\bth}\,$ with 
$\,I=3$.
The mean of this fitted distribution is 22.75.  
The distribution in Fig.~\ref{BCBH.fig} appears as a mixture of a distribution concentrated near $\,k=0\,$ and a left-truncated normal distribution with a local mode at 24.
The observed value BH=29 is indicated in this figure.

The point mass at BH=0 is about 0.1 and values of $\,{\rm BH}\leq 3\,$ account for about 20\% of the distribution with $\,I=3\,$ and fitted $\,\widehat{\bth}$. 
In Appendix~B we show this distribution is approximately a mixture of a Borel-Tanner distribution near zero and a normal with mean 26.1 and standard deviation of 14.9.

\subsection{The Cancer Genome Atlas: Lung Cancer}
\label{lung.section}

This dataset contains the summary of an extensive database collected on tumors from 178 patients with squamous cell lung carcinoma. 
A full description of these data and the analyses performed are summarized in the Cancer Genome Atlas~(2012).\label{TCGA.2012} 
The data values were downloaded from the website
{\tt https://tcga-data.nci.nih.gov/}.
We choose to examine p-values representing summaries of statistical comparisons of smokers and non-smokers across the genetic markers. 
We identified $\,n = 20,068\,$ observed p-values after omitting about 2\% missing values.

\begin{figure}[t]
\vspace*{-.45in}
\begin{center}
\includegraphics[scale=.7]{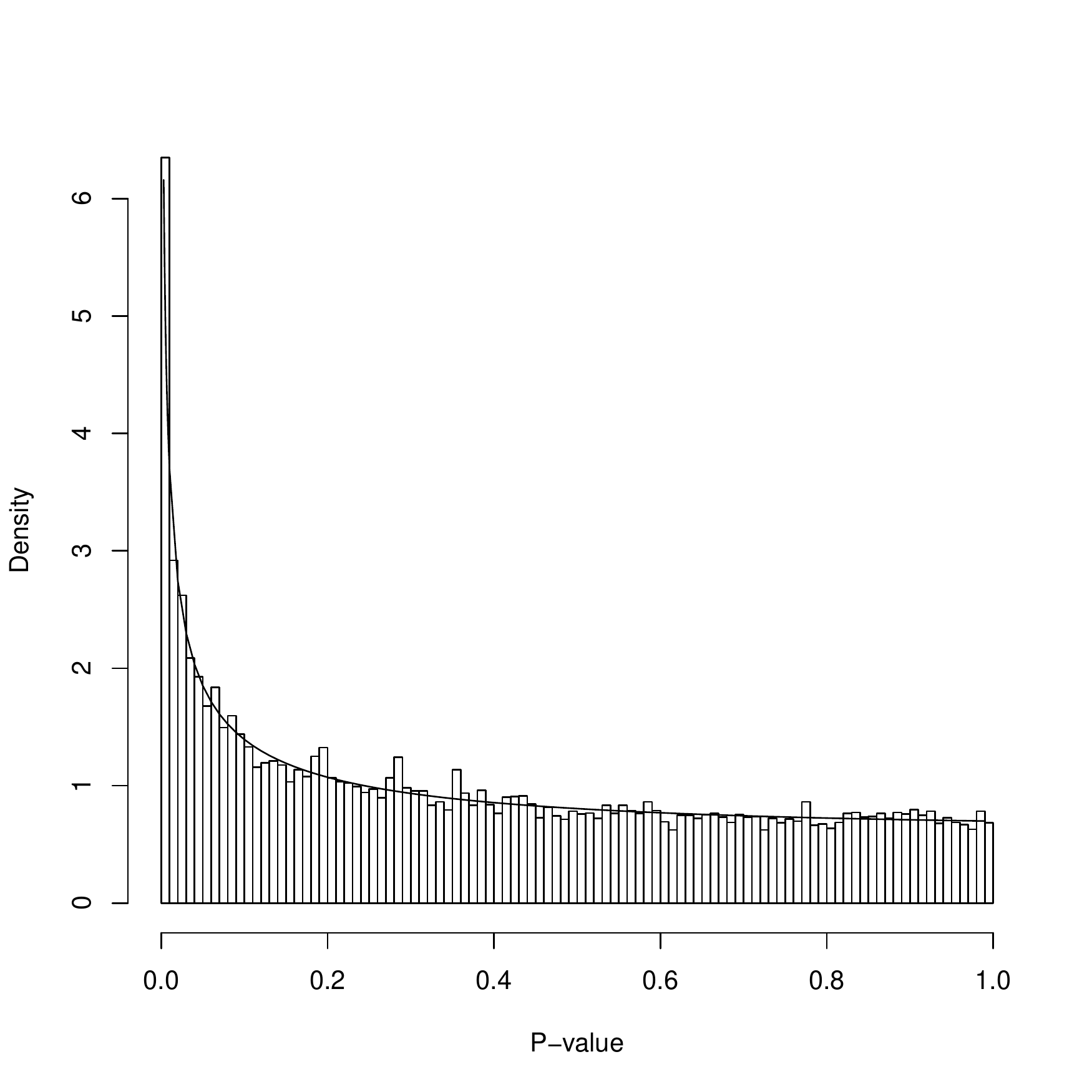}
\end{center}
\vspace*{-.35in}
\caption{Observed and fitted $\,\psi_4\,$ for 20,068 TCGA lung cancer p-values.  \label{AtlasPvalue.fig}}
\vspace*{.35in}
\end{figure}

Using the BH procedure, 173 p-values are identified with FDR $\,=.05$.
The fitted parameter values $\,\widehat{\bth}\,$ are given in Table~\ref{fitted.table}.  
Distributions up to $\,I=4\,$ showed statistically significant improvement in the log-likelihood but larger values of $\,I\,$ failed to change it.
The fitted density function $\,\psi_4(\cdot\mid\widehat{\bth})\,$ appears in Fig.~\ref{AtlasPvalue.fig} and demonstrates good agreement with the observed data.
The estimate $\,\widehat\theta_0\,$ of $\,\pi_0\,$ is about .70 for $\,I=4$.

\typeout{}\typeout{TCGA observed and fitted p-values}\typeout{}
\typeout{}\typeout{Atlas fitted BH model}\typeout{}
\begin{figure}[t]
\vspace*{-.35in}
\begin{center}
\includegraphics[scale=.7]{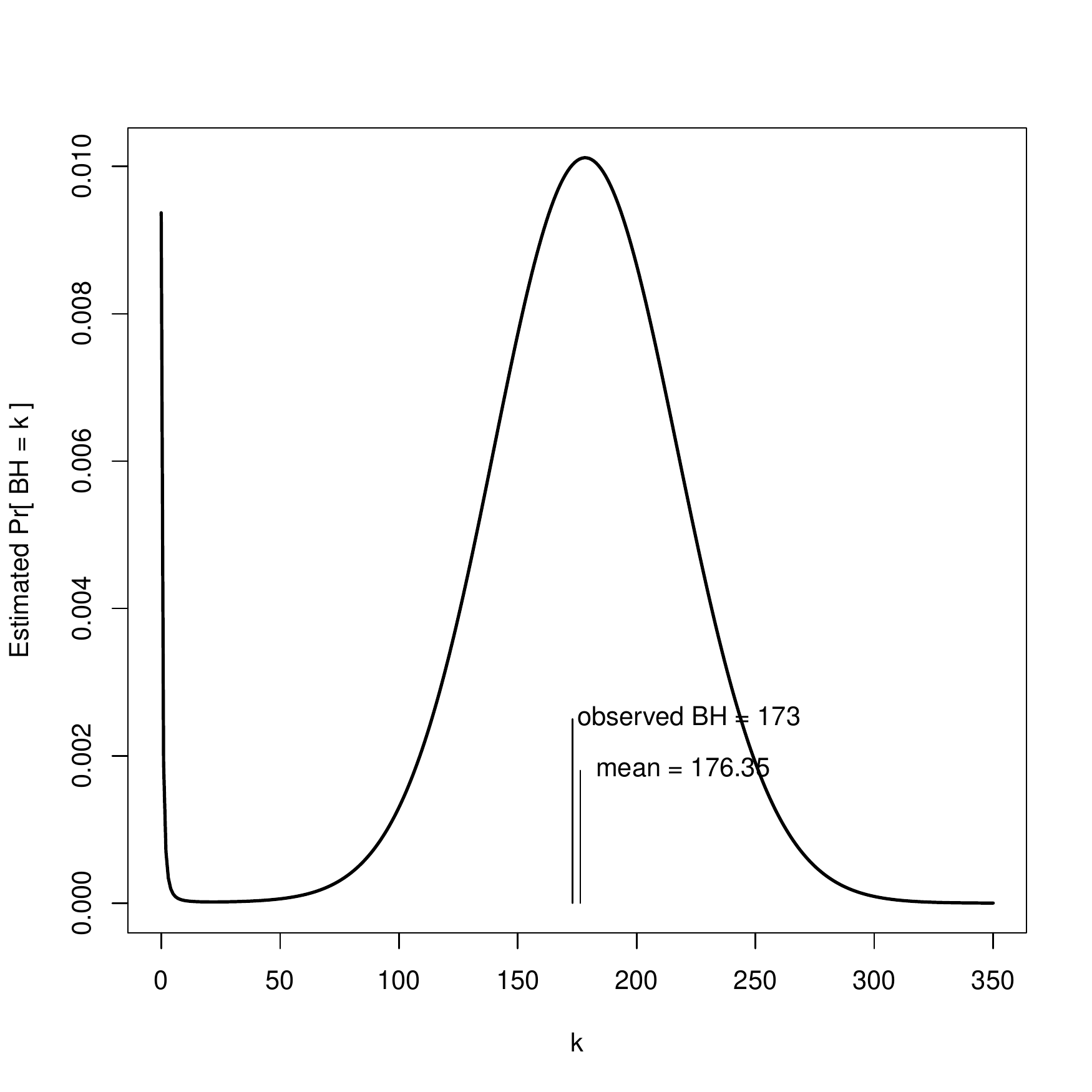}
\end{center}
\vspace*{-.35in}
\caption{Fitted BH distribution for TCGA lung cancer p-values.  \label{AtlasBH.fig}}
\vspace*{.35in}
\end{figure}

The fitted BH distribution~(\ref{BH.eq}) is plotted in 
Fig.~\ref{AtlasBH.fig}. 
There is close agreement between the observed value (173), the mean (176.35) of the fitted distribution, and the local mode (177).
As with Fig.~\ref{BCBH.fig}, the fitted distribution of BH appears as a mixture of a distribution concentrated near zero and a normal  distribution.
The local mode at zero gives a fitted $\,\Pr[\,{\rm BH}\leq 2\,]\,$ of    .012.

As in Fig.~\ref{BCBH.fig}, the probability mass near zero corresponds to the Borel-Tanner distribution~(\ref{Borel.eq}). 
The density mass away from zero is approximately that of a normal distribution with mean 178.8 and standard deviation 39.1.
Details are given in Appendix~B.

\section{Sampling Dependent P-values}
\label{dependence.section}
\typeout{}
\typeout{Section \thesection: Sampling Dependent P-values}
\typeout{}

In this section we create two different methods for describing sampling of dependent p-values: one based on the distribution of a single order statistic and a second method conditioning on an unobservable, latent variable.
In both cases, greater dependence among the p-values results in greater means and variances for the distribution of p-values identified by Bonferroni and BH methods.  
These behaviors are also described by Owen~(2005).\label{Owen.2005.1}
Greater dependence also contributes to a larger point mass at zero.
We will use the fitted breast cancer example of Section~\ref{breast.cancer.section} to illustrate these methods.

\subsection{Order statistics from an exchangeable parent}
\label{MM.section}
\typeout{}
\typeout{Section \thesubsection: Order statistics from M\&M}
\typeout{}

Let us assume the p-values are marginally sampled from the fitted distribution $\,\Psi_3(\,\cdot\mid\widehat{\bth})\,$ of the breast cancer example in Section~\ref{breast.cancer.section} with $\,n=3226$. 
The probability of finding a specified p-value identified as statistically significant with FDR$\,\leq\alpha=.05\,$ using the Bonferroni correction of $\,\alpha/n=1.55\times 10^{-5}\,$ will then occur with probability
$$
    p^* = \Psi_3(\alpha/n\mid\widehat{\bth}) = 7.12\times 10^{-4} \, .
$$

Let \,B\, denote the number of p-values identified using the Bonferroni correction defined at~(\ref{B.def.eq}) with FDR$\,\leq.05$.
Then $\,p_{\rm B}\leq\alpha/n\,$ and $\,p_{\rm B+1}>\alpha/n\,$ so
$$
      \Pr[\,{\rm B}\geq k\,] = \Pr[\,p_k \leq p^*\,]
$$
for $\,k=0, 1, \ldots\,$ with ordered values $\,p_1\leq\cdots\leq p_n\,$.

Let $\,C(u_1,\ldots,u_m)\,$ denote the joint, exchangeable, cummulative distribution function of any set of size $\,m\,$ p-values.
From Maurer and Margolin~(1976, eqn (1.2)),\label{Maurer.1976} we then have
$$
  \Pr[\,p_k \leq p^*\,] \; = \;\sum_{m=k}^n \;
      {{m-1}\choose{k-1}} \,{{n}\choose{m}}
        C(p^*, \ldots , p^*)  
$$
where the argument to $\,C\,$ contains $\,m\,$ copies of $\,p^*$.

Specifically we chose to model dependence among the p-values using a Gumbel exchangeable copula model with joint cumulative distribution function
$$
     C(u_1, \ldots , u_m\mid\gamma) = \exp\left\{
              -\left(\sum\; (-\log u_i)^\gamma
                  \right)^{1/\gamma}\right\}
$$
for $\,0<u_i\leq 1\,$ and parameter $\,\gamma\geq 1\,$ controlling the degree of dependence so that
$$
       C(p^*,\ldots, p^*) = (p^*)^{m^{1/\gamma}}\; .
$$

\typeout{}
\typeout{Dependent Bonferroni sampling}\typeout{}
\begin{figure}[t]
\vspace*{-.35in}
\begin{center}
\includegraphics[scale=.7]{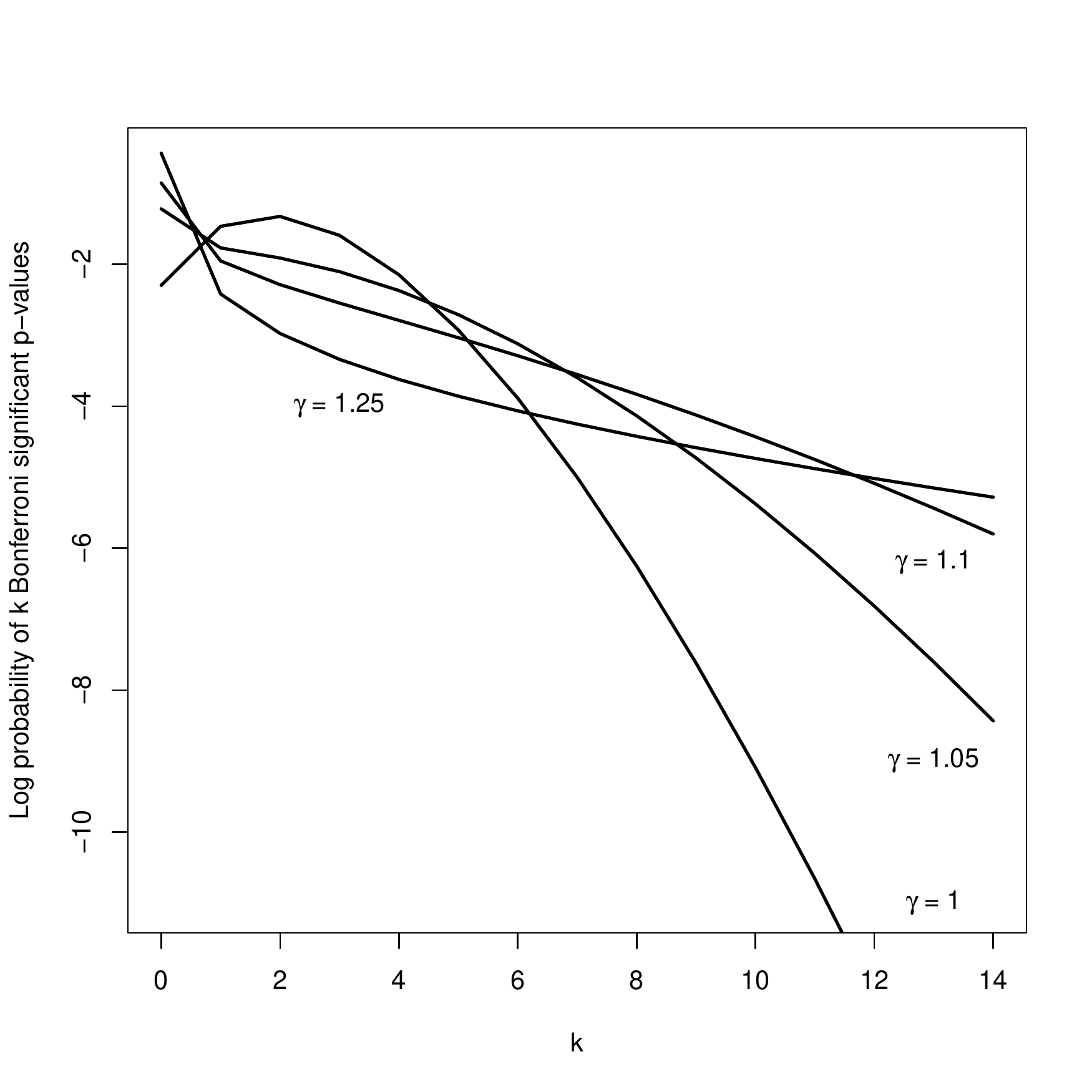}
\end{center}
\vspace*{-.35in}
\caption{Log probability of $\,k\,$ p-values identified using the Bonferroni method, marginally sampled from the fitted parameters in the breast cancer example with FDR$\,\leq.05$. 
The p-values are jointly sampled from a Gumbel copula distribution with parameters $\,\gamma\,$ as given. \label{depBonf.fig}}
\vspace*{.35in}
\end{figure}

We computed this probability and plot $\,\log\Pr[{\rm B}=k]\,$ for values of $\,\gamma\,$ given in Fig.~\ref{depBonf.fig}.
The Poisson distribution with mean
$$
     n\Psi_3(\alpha/n\mid\widehat{\bth}) = np^* = 2.297
$$
coincides with the independence model for $\,\gamma=1$.
The upper tails of these distributions increase with the greater dependence corresponding to larger values of $\,\gamma$. 
Similarly, larger point masses $\,\Pr[\,{\rm BH}=0\,]\,$ are also associated with greater dependence.

\subsection{A latent variable approach}
\label{latent.section}
\typeout{}
\typeout{Section \thesubsection: A latent vairable approach}
\typeout{}

Let $\,\bth\,$ and $\,\beps\,$ denote $\,I-$tuples such that both $\,\bth+\beps\,$ and $\,\bth-\beps\,$ are valid parameters for the distribution $\,\Psi_I\,$ described in Section~\ref{Psi.section}.
Let $\,Y\,$ denote a Bernoulli random variable with parameter equal to 1/2.
Conditional on the (unobservable) value of $\,Y,$ assume all p-values are sampled from either $\,\Psi_I(\,\cdot\mid\bth+\beps)\,$ or $\,\Psi_I(\,\cdot\mid\bth-\beps)$.
The marginal distribution of these exchangeable p-values is then $\,\Psi_I(\cdot\mid\bth)\,$ using~(\ref{mixture.eq}).

To demonstrate the correlation among the p-values induced by this latent model, let $\,Q_1,\,Q_2\,$ denote a random sample from $\,\Psi_I,$ both with parameters either $\,\bth+\beps\,$ or $\,\bth-\beps,$ conditional on $\,Y$.
The $\,Q_i\,$ are conditionally independent given $\,Y\,$ and have marginal covariance
$$
  {\rm Cov}(Q_1,\, Q_2) \;=\; \{\mu(\bth+\beps)\}^2 /2 
            \; +\; \{\mu(\bth-\beps)\}^2/2 \; -\;\{\mu(\bth)\}^2
$$
where $\,\mu(\bth)\,$ is the mean of $\,\psi_I(\,\cdot\mid\bth)\,$ calculated from~(\ref{Psi.moment.eq}).
This covariance is always positive.

Continuing to sample in this fashion, we then have
\begin{equation}         \label{BH.dependent.eq} 
   \Pr[\,{\rm BH} \; = \; k\mid\bth\,] \; =\; 
         \Pr[\,{\rm BH} = k\mid\bth-\beps\,]\, /2 \; + \;
         \Pr[\,{\rm BH} = k\mid\bth+\beps\,]\, /2
\end{equation}
and can evaluate this expression using~(\ref{BH.eq}).

\begin{figure}[t]
\vspace*{-.35in}
\begin{center}
\includegraphics[scale=.7]{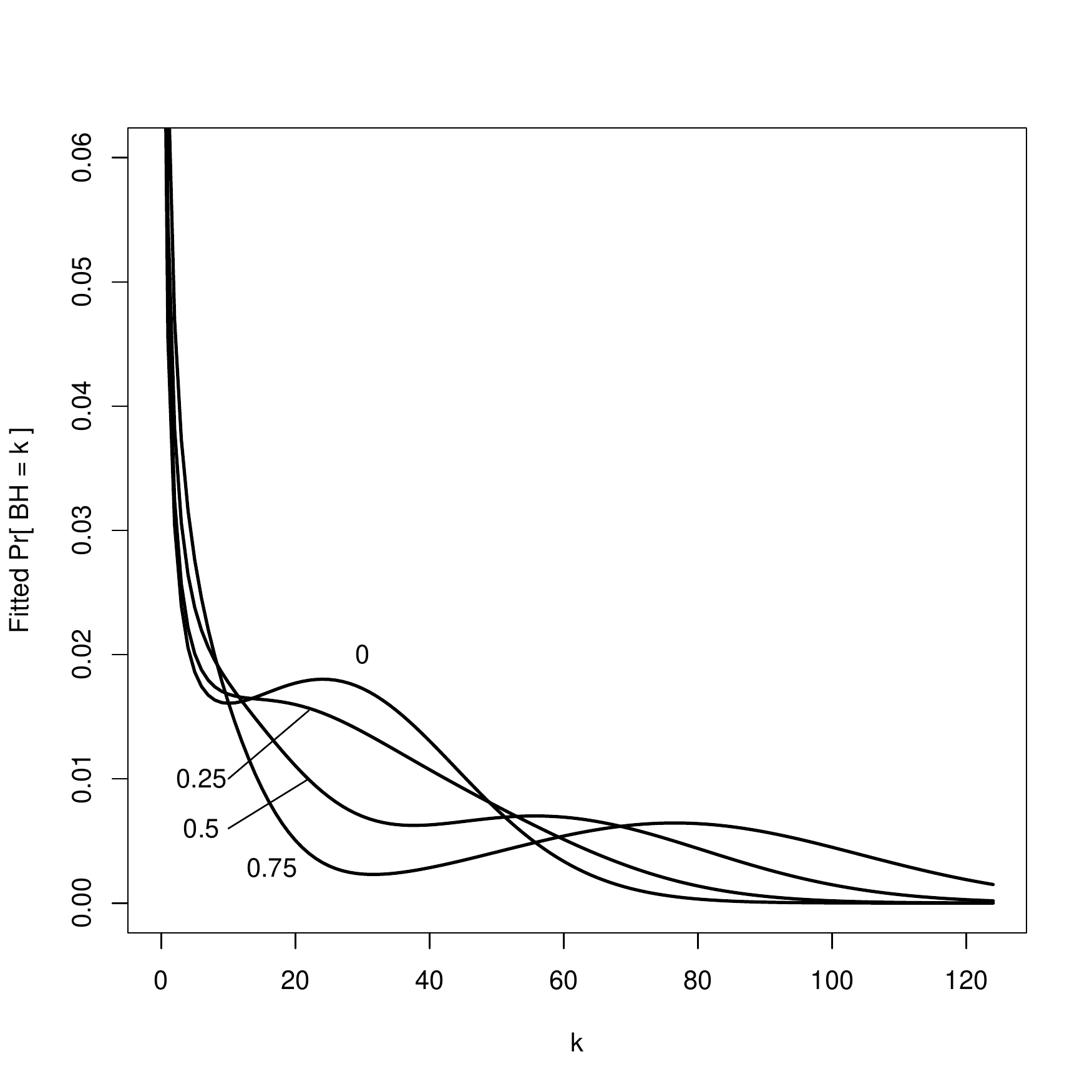}
\end{center}
\vspace*{-.35in}
\caption{Distributions of BH under dependent sampling for fitted parameters from the breast cancer example. 
Values of $\,z\,$ are given and control the dependence among the underlying p-values.
Summaries of these distributions are given in Table~\ref{BH.dependent.table}.\label{dependent.fig}}
\vspace*{.35in}
\end{figure}

As an illustration, we used $\bth = \widehat{\bth}\,$ and $\,\beps = z \widehat{\bsig}\,$ where $\,\widehat{\bth}\,$ and $\,\widehat{\bsig}\,$ are the fitted parameters and their estimated standard errors respectively given in Table~\ref{fitted.table} for the breast cancer example with $\,I=3$.
The distributions given at~(\ref{BH.dependent.eq}) for $\,z=0, $ .25, .5, and .75 are plotted in Fig.~\ref{dependent.fig}. 
Summaries of these four distributions and the mutual correlations of the p-values are given in Table~\ref{BH.dependent.table}.
As with Figs.~\ref{BCBH.fig} and~\ref{AtlasBH.fig}, all distributions in Fig.~\ref{dependent.fig} appear as mixtures of distributions concentrated near zero and a truncated normal distributions away from zero.
Greater dependence results in a larger point mass at zero, as well as larger means and variances of BH. 
Increased variances in this setting are also described by Owen~(2005).\label{Owen.2005.2}

\typeout{}
\typeout{Dependent BH sampling}\typeout{}

\begin{table}[t]    
\begin{center}
\caption{Properties of the BH distributions sampling correlated p-values using~(\ref{BH.dependent.eq}). 
These distributions are plotted in Fig.~\ref{BH.dependent.table}. 
\label{BH.dependent.table}}
\begin{tabular}{ccccc} \\ \hline
    & Correlation \\
$z$ & of p-values & Mean & SD & $\Pr[\,{\rm BH} = 0\,]$ \\ \hline
 0\p  &  0\q  & 22.75 & 18.13 & .101 \\
 .25  & .004  & 24.43 & 21.44 & .104 \\
 .5\p & .017  & 29.40 & 29.50 & .116 \\
 .75  & .037  & 37.18 & 39.85 & .136\\ \hline
\end{tabular}
\end{center}
\end{table}

\section{Power for Planning Studies}
\label{power.section}
\typeout{}
\typeout{Section \thesection: Power for planning studies}
\typeout{}

In this final section we describe how to plan for a larger project using  data from a smaller pilot study.
Huang, Wu, Su, {\it et al}~(2015)\label{Huang.2015.1} report on a study of $\,N=78\,$ patients with lung cancer and examined $\,n=48,803\,$ markers to determine if any of these are related to patient survival. 
(A link to their data appears in the References.)
None of these markers were identified as statistically significant at FDR$\,=.05$ using the Bonferroni method.

We examined their data and the parameter estimates for our fitted models  $\,\psi_I\,$ appear in Table~\ref{fitted.table}.
We found the model with $\,I=3\,$ provided the best fit and worked with the maximum likelihood estimated $\,\widehat{\bth}\,$ to model power.
We estimate over 90\% of the p-values were sampled from the null hypothsis in these data.

In order to describe power we will assume the magnitude of the effect, as measured by $\,\bth,\,$ is proportional to the square root of the subject sample size, as is often the case with parameters whose estimates are normally distributed. 
This assumption will also require values of $\,\bth\,$ to lie near the center of the valid parameter space.

\begin{table}[t]  
\begin{center}
\caption{Estimated power based on pilot data from Huang {\it et al.}~(2015).\label{power.table}}
\begin{tabular}{ccc c cc} \\ \hline
 Sample & \multicolumn{2}{c}{Dependence\quad} & &
 \multicolumn{2}{c}{Estimated} \\ \cline{2-3}\cline{5-6}
 size & $z$  & Correlation && Expected BH & $\Pr[{\rm BH}>0]$ \\ \hline
 \p78 & 0   & 0    & & \p1.5 & 0.517 \\
     & 0.4 & .001 &      & \p1.7 & 0.499 \\
     & 0.8 & .006 &      & \p2.7 & 0.444 \\[2ex]
 300 & 0   & 0    & & \p6.5 & 0.748 \\
     &0.4 & .006 &      &  11.4 & 0.712 \\
     & 0.8 & .002 &      &  30.9 & 0.592\\[2ex]
 450 &  0   & 0    & &  12.6 & 0.813 \\
     & 0.4 & .008 &      &  26.2 & 0.772 \\
     & 0.8 & .034 &      &  75.0 & 0.631 \\[2ex]
 600 & 0   &    0 &  &  21.7 & 0.855 \\
     & 0.4 & .011 &      &  49.0 & 0.812 \\
     & 0.8 & .045 &      &  90.8 & 0.657 \\ \hline
\end{tabular}
\end{center}
\end{table}

Let $\,\widehat{\bth}\,$ denote the maximum likelihood estimate in $\,\psi_3\,$ for the Huang {\it et al}~(2015) data given in Table~\ref{fitted.table}.
We computed power estimates in Table~\ref{power.table} setting 
$$
          \bth =\bth(N) = (N/78)^{1/2}\;\widehat{\bth}
$$
where $\,N\,$ is the proposed sample size and used $\,\beps = z\bth\,$ in~(\ref{BH.dependent.eq}) to vary the dependence among p-values for values of $\,z=0, .4,$ and .8.

A variety of sample sizes and correlations are summarized in Table~\ref{power.table}.
This table summarizes the power as the probability of identifying at least one marker with FDR$\,=.05$.
The expected number of identified findings using BH is also given in this table.

We estimate the published study by Huang {\it et al}~(2015)\label{Huang.2015.2} had about a 50\% chance of detecting at least one marker with FDR$\,=.05$.
Table~\ref{power.table} shows sample sizes of $\,N\geq450$ would have power greater than 80\% under a model of independent sampling.
Even small mutual correlations result in greater point masses at zero, reducing the power of detecting at least one statistically significant p-value.


\section*{References}
\begin{description}

\item
Benjamini Y. (2010). Discovering the false discovery rate. {\it Journal of the Royal Statistical Society} B {\bf 72}: 405--16. [\pageref{Benjamini.2010}].

\item 
Benjamini Y, and Hochberg Y (1995). Controlling the false discovery rate: A practical and powerful approach to multiple testing. {\it Journal of the Royal Statistical Society} B {\bf 57}: 289--300.
[\pageref{BH.1995.1}, \pageref{BH.1995.2}].

\item
Benjamini Y, and Hochberg Y (2000). On the adaptive control of the false discovery rate in multiple testing with independent statistics. {\it Journal of Educational and Behavioral Statistics\/} {\bf 25.1}: 60--83.[\pageref{BH.2000}].

\item
Broberg P (2005). A comparative review of estimates of the proportion unchanged genes and the false discovery rate. {\it BMC Bioinformatics} {\bf 6}: 199--218.  [\pageref{Broberg.2005.1}, \pageref{Broberg.2005.2}].

\item 
Cancer Genome Atlas Research Network (2012). Comprehensive genomic characterization of squamous cell lung cancers. {\it Nature\/} {\bf 489}: 519--25.  [\pageref{TCGA.2012}]. 

\item
Donoho D and Jin J (2004). Higher criticism for detecting sparse heterogeneous mixtures. {\it Annals of Statistics} {\bf 32}: 962--94.
 [\pageref{Donoho.2004}]. 

\item 
Efron B, Tibshirani R, Storey JD, Tusher V (2001).
Empirical Bayes analysis of a microarray experiment. 
{\it Journal of the American Statistical Association\/} {\bf 96}:1151--60.
 [\pageref{Efron.2001}].  

\item 
Efron B (2004). Large-scale simultaneous hypothesis testing: The choice of a null hypothesis. {\it Journal of the American Statistical Association} {\bf 99}: 96--104. 
 [\pageref{Efron.2004.1}, \pageref{Efron.2004.2}]. 

\item
Friguet C, Kloareg M, and Causeur D (2009). A factor model approach to multiple testing under dependence. {\it Journal of the American Statistical Association} {\bf 104}:  1406--15.
 [\pageref{Friguet.2009}]. 

\item
Genovese C, and Wasserman L (2004). A stochastic process approach to false discovery control. {\em Annals of Statistics\/} {\bf 32}: 1035--61.
 [\pageref{Genovese.2004.1}, \pageref{Genovese.2004.2}]. 

\item
Hedenfalk I, Duggan D, Chen Y, {\em et al.} (2001). Gene-expression profiles in hereditary breast cancer. {\it New England Journal of Medicine} {\bf 344}: 539--48. [\pageref{Hedenfalk.2001}].

\item
Huang H-L, Wu Y-C, Su L-J, {\it et al.} (2015). Discovery of prognostic biomarkers for predicting lung cancer metastasis using microarray and
survival data. {\it BMC Bioinformatics} {\bf 16}:54  [\pageref{Huang.2015.1}, \pageref{Huang.2015.2}].
 Their data is available at \\
{\tt www.biomedcentral.com/content/supplementary/s12859-015-0463-x-s1.xls} 

\item 
Jin J, and Cai TT (2007). Estimating the null and the proportion of nonnull effects in large-scale multiple comparisons. {\it Journal of the American Statistical Association} {\bf 102}: 495--506.
 [\pageref{Jin.2007.1}, \pageref{Jin.2007.2}].

\item
Jolley LBW (1961). {\it Summation of Series.} Second edition.  New York: Dover.  [\pageref{Jolley.1961}]. 

\item
Kozoil JA, and Tuckwell HC (1999). A Bayesian method for combining statistical tests. {\it Journal of Statistical Planning and Inference\/} {\bf 78}: 317--23.  [\pageref{Kozoil.1999}].

\item 
Langaas M, Lindqvist BH, and Ferkingstad E (2005).  Estimating the proportion of true null hypotheses, with application to DNA microarray data. {\it Journal of the Royal Statistical Society\/} B  {\bf 67}: 555--72.  [\pageref{Langaas.2005.1}, \pageref{Langaas.2005.3}].

\item
Maurer W, and Margolin BH (1976). The multivariate inclusion-exclusion formula and order statistics from dependent variates. {\it Annals of Statistics\/} {\bf 4}: 1190--9. [\pageref{Maurer.1976}].

\item
Owen AB (2005). Variance of the number of false discoveries. {\it Journal of the Royal Statistical Society}: Series B, {\bf 67}: 411--26. [\pageref{Owen.2005.1}, \pageref{Owen.2005.2}].

\item
Pounds S, and Morris SW (2003).  Estimating the occurrence of false positives and false negatives in microarray studies by approximating and partitioning the empirical distribution of p-values. {\it Bioinformatics.\/} {\bf 19}: 1236--42. [\pageref{Pounds.2003.1}, \pageref{Pounds.2003.2}].

\item
Ruiz SM (1996). An algebraic identity leading to Wilson's Theorem. {\em The Mathematical Gazette\/} 80.{\bf 489}: 579--82. [\pageref{Ruiz.1996}].

\item
Storey JD, and Tibshirani R (2003). Statistical significance for genomewide studies. {\it Proceedings of the National Academy of Science U S A} {\bf 100}: 9440-5.  [\pageref{Storey.2003}].

\item
Sun W, and Cai TT (2009).  Large-scale multiple testing under dependence. {\it Journal of the Royal Statistical Society} Series B {\bf 71}: 393--424. [\pageref{Sun.2009}].

\item
Tang Y, Ghosai S, and Roy A (2007).  Nonparametric Bayesian estimation of positive false discovery rates. {\it Biometrics\/} {\bf 63}: 1126--34. 
[\pageref{Tang.2007.1}, \pageref{Tang.2007.2}, \pageref{Tang.2007.3}].

\item
Tanner JC (1961). A derivation of the Borel distribution. 
{\it Biometrika\/} {\bf 48}: 222--4.  [\pageref{Tanner.1961}].

\item
Wu W (2008). On false discovery control under dependence. {\it Annals of Statistics} {\bf 36}: 364--80. [\pageref{Wu.2008}].

\item
Yu C, and Zelterman D. (2017).  A parametric model to estimate the proportion from true null using a distribution for {\it p-}values.
{\it Computational Statistics \& Data Analysis\/} {\bf 114}: 105--18.
 [\pageref{Yu.2017}].

\end{description}


\section*{Appendix A: Details of 
Section~\ref{basic.results.section}}

At~(\ref{Uk.eq}) we define $\,U_0=1\,$ and 
$$
 U_k = \int_{p_1=0}^{\alpha/n}
       \int_{p_2=p_1}^{2\alpha/n}
                 \;\cdots\;
       \int_{p_k=p_{k-1}}^{k\alpha/n}   
       \psi(p_1)\,\cdots\,\psi (p_k)\, 
       {\rm d}p_k\,\ldots\,{\rm d}p_2\,{\rm d}p_1 
$$
for $\,k=1,2,\ldots$.

To demonstrate~(\ref{Uk.recursive.eq}), we integrate one term at a time to show
\begin{eqnarray*}
   U_k &=& \int_{p_1=0}^{\alpha/n}
       \int_{p_2=p_1}^{2\alpha/n}
                 \;\cdots\;
       \int_{p_{k-1}=p_{k-2}}^{(k-1)\alpha/n}\;\; 
       \{\Psi(k\alpha/n) - \Psi(p_{k-1})\}\,   
       \psi(p_1) \cdots \psi (p_{k-1})\, 
       {\rm d}p_{k-1}\cdots {\rm d}p_2 \,{\rm d}p_1 \\
    &=&
    \Psi(k\alpha/n)\,U_{k-1}\, -
    \int_{p_1=0}^{\alpha/n}
    \int_{p_2=p_1}^{2\alpha/n}
                 \;\cdots\;
       \int_{p_{k-2}=p_{k-3}}^{(k-2)\alpha/n} \;
       \{\Psi^2((k-1)\alpha/n) - \Psi^2(p_{k-2})\}/2!   \\
       && \hspace*{.25in}
       \times\psi(p_1) \,\cdots \,\psi (p_{k-2}) \,
       {\rm d}p_{k-2}\ldots {\rm d}p_2 \,{\rm d}p_1 \\
   &=&
    \Psi(k\alpha/n)\,U_{k-1} - \Psi^2\{(k-1)\alpha/n\}\,U_{k-2}/2! \\
    &&  \hspace*{.25in}
    +\frac{1}{2!} \int_{p_1=0}^{\alpha/n}
    \int_{p_2=p_1}^{2\alpha/n}
                 \;\cdots\;
       \int_{p_{k-2}=p_{k-3}}^{(k-2)\alpha/n}\; 
       \Psi^2(p_{k-2})\,
       \psi(p_1)\,\cdots\,\psi(p_{k-2})\, 
       {\rm d}p_{k-2}\,\ldots\,{\rm d}p_2\,{\rm d}p_1 
\end{eqnarray*}
\noindent and continue in this manner to demonstrate the recursive relation
\begin{equation}                    
    U_k = \sum_{i=1}^k \; (-1)^{i+1} \,    \label{A.Uk.recursive.eq}
           \Psi^i\{(k-i+1)\alpha/n\}\, U_{k-i}/\,i!
\end{equation}
given at~(\ref{Uk.recursive.eq}).

To demonstrate~(\ref{exact.eq}) for the specific case of $\,\Psi(p) = p\,$ we need to show
\begin{equation}                           
  U_k = (k+1)^{k-1}(\alpha/n)^k\, /\, k! \; . \label{Uk.uniform.eq}
\end{equation}

We will prove~(\ref{Uk.uniform.eq}) by induction on $\,k$.

In Section~\ref{basic.results.section} we demonstrate~(\ref{Uk.uniform.eq}) is true for $\,k=0, 1, 2$.
Next, we assume if~(\ref{Uk.uniform.eq}) is valid for any $\,k=0,1,\ldots\,$ then it is also true for $\,k + 1$.

Begin by using the recursive relation~(\ref{A.Uk.recursive.eq}) with $\,\Psi(p) = p\,$ and~(\ref{Uk.uniform.eq}) for $\,k\,$ giving
\begin{eqnarray*}
   U_{k+1} &=& \sum_{i=1}^{k+1} \; (-1)^{i+1}
        \left\{\frac{(k-i+2)\alpha}{n}\right\}^i
        \left\{\frac{(k-i+2)^{k-i}\alpha^{k-i+1}}
                    {(k-i+1)! \, i! \, n^{k-i+1}}  \right\}\\
    &=& (\alpha/n)^{k+1}\;\sum_{i=1}^{k+1} \; (-1)^{i+1}
        \frac{(k-i+2)^k}{(k-i+1)!\, i!} 
\end{eqnarray*}

It remains to show
$$
   \sum_{i=1}^{k+1}\; (-1)^{i+1} (k-i+2)^k / (k-i+1)!\, i!
      \; = \; (k+2)^k / (k+1)!
$$
or equivalently
$$
   \sum_{i=0}^{k+1}\; (-1)^{i+1}{{k+1}\choose{i}} (k-i+2)^k = 0\, .
$$

Continue by writing $\,{{k+1}\choose{i}} = {{k}\choose{i}} + {{k}\choose{i-1}}\,$ and set $\,j=i-1\,$ giving
\begin{eqnarray*}
   \sum_{i=0}^{k+1}\; (-1)^{i+1}{{k+1}\choose{i}} (k-i+2)^k &=& 
   \sum_{i=0}^k\; (-1)^{i+1}{{k}\choose{i}} (k-i+2)^k  \\
   &&\quad +\;\sum_{j=0}^k\; (-1)^j{{k}\choose{j}} (k-j+1)^k \, .
\end{eqnarray*}

The proof of~(\ref{Uk.uniform.eq}) is completed by two applications of the Ruiz Identity (Ruiz, 1996).\label{Ruiz.1996}
Specifically,
$$
           \sum_{i=0}^k\; (-1)^i {{k}\choose{i}} (x-i)^k = k!
$$
for all integers $\,k\geq 0\,$ and all real numbers $\,x$.

\section*{Appendix B: Asymptotic, non-null distributions}

Here we demonstrate the distribution of B and BH when a large number of p-values are independently sampled from $\,\Psi_I(p\mid\bbeta)\,$ for $\,I\geq 1$. 
The mixture of distributions for BH is readily apparent in Figs.~\ref{BCBH.fig} and~\ref{AtlasBH.fig}.
Briefly, we obtain either a Borel or a normal limiting distribution of BH  depending on $\,\bbeta$.
Similarly, the limiting distribution of the number of p-values identified by the Bonferroni method can be either Poisson or normal.

To describe the behavior of B and BH for values of $\,k\,$ near zero and large values of $\,n$, consider a sequence of parameter values $\,\bbeta_n=\bbeta/(\log n)^I\,$ shrinking to zero with $\,I\geq 1$.
Following~(\ref{Psi.eq}), we always have $\,\beta_0=1$.

Begin by writing
\begin{eqnarray}
  n\Psi_I(\gamma/n\mid \bbeta_n) &=& 
  \gamma\left\{ 1\; + \;\frac{\beta_1}{(\log n)^I} (\log n - \log\gamma) 
           \; +\cdots+\;\;    
   \frac{\beta_I}{(\log n)^I} (\log n - \log\gamma)^I  \right\} 
       \nonumber \\
   &=& \gamma(\beta_I +1) + O(1/\log n)           \label{asymP.eq}
\end{eqnarray}                             
for any fixed $\,\gamma>0$.

When sampling from $\,\Psi_I(\cdot\mid\bbeta_n)\,$  using the Bonferroni rule~(\ref{B.def.eq}), set $\,\gamma=\alpha\,$ in~(\ref{asymP.eq}) to demonstrate the number of statistically significant p-values B will have an approximate Poisson distribution with mean $\,\alpha(\beta_I +1)$.

Similarly, under the parameter sequence $\,\bbeta_n\,$ the distribution of BH near zero will be approximated by the Borel distribution~(\ref{Borel.eq}) with parameter $\,\alpha(\beta_I+1)$.
More formally, we will show if $\,n\,$ p-values are independently sampled from 
$\,\Psi_I(\,\cdot\mid\bbeta/(\log n)^I)\,$ then
\begin{equation}    \label{Approx.Borel.eq}     
    \Pr[\,{\rm BH} = k \,] = 
    (k+1)^{k-1}/k!\;\; \{\alpha(\beta_I+1)\}^k\, 
    \exp\{-(k+1)\alpha(\beta_I+1)\}  \; + O(1/\log n)
\end{equation}
for moderate values of $\,k=0,1,\ldots$.

Following~(\ref{asymP.eq}), we have
$$
   \{1-\Psi_I((k+1)\alpha/n\mid\bbeta_n)\}^{n-k} 
   = \exp\{-(k+1)\alpha(\beta_I + 1)\} \; + O(1/\log n)
$$
demonstrating~(\ref{Approx.Borel.eq}) for 
\begin{eqnarray*}
   \Pr[\,{\rm BH} = 0\mid\bbeta_n\,] &=&
        \exp\{-\alpha(\beta_I+1)\} \; + O(1/\log n)\; ,  \\
    \Pr[\,{\rm BH} = 1\mid\bbeta_n\,] &=& \alpha(\beta_I+1)\, 
            \exp\{ -2\alpha(\beta_I+1)\} \; + O(1/\log n)\; ,
\end{eqnarray*}
\noindent 
and
$$
\Pr[\,{\rm BH}=2\mid\bbeta_n\,] = 3/2\;\{\alpha(\beta_I+1)\}^2\, 
            \exp\{ -3\alpha(\beta_I+1)\} \; + O(1/\log n)\; .
$$

The remainder of the proof of~(\ref{Approx.Borel.eq}) for other values of $\,k\,$ closely follows the proof by induction of~(\ref{Uk.uniform.eq}) in Appendix~A to show
$$
   n^kU_k(\bbeta_n) = (k+1)^{k-1}\{\alpha(\beta_I+1)\}^k/k! 
        \; +  O(1/\log n)\; .    
$$

For a fixed value of $\,\bbeta\,$ and large $\,n,$ the Bonferroni B will behave approximately as Poisson with mean 
$\,n\Psi_I(\alpha/n\mid\bbeta)$.  
In the example of the fitted lung cancer data of Section~\ref{lung.section}, this value is $\,n\Psi_4( \alpha /n\mid\widehat\bbeta)= 4.67$.

To describe the approximate behavior of BH away from zero, consider the fitted quantile function $\,\Psi_I^{-1}(i/n\mid\widehat\bbeta)\,$ giving the approximate expected value of the order statistic $\,p_i$.
BH is the smallest value of $\,k\,$ for which 
$\,p_{k+1} > (k+1)\alpha/n.$ 
This should occur for values of BH with mean $\,\mu= \mu(\widehat\bbeta)\,$ solving
$$
      \Psi_I^{-1}((\mu+1) / n\mid\widehat\bbeta) = (\mu+1)\alpha/n
$$
or equivalently,
$$
           \Psi_I((\mu+1)\alpha/n\mid\widehat\bbeta) = (\mu+1)/n
$$
giving estimated values $\,\mu(\widehat\bbeta) = 26.1\,$ for the breast cancer example in Fig.~\ref{BCBH.fig} and $\,\mu(\widehat\bbeta) = 178.8\,$ for the TCGA lung cancer example in Fig.~\ref{AtlasBH.fig}.

The approximate standard deviation of BH is
$$
     \sigma(\widehat\bbeta) = 
     \left\{
       \,n\left/\,\psi_I(\mu(\widehat\bbeta)\alpha/n \mid \widehat\bbeta)
     \right.\right\}^{1/2}
$$
giving $\,\sigma(\widehat\bbeta) =14.9\,$ for the breast cancer example and $\,\sigma(\widehat\bbeta) = 39.1\,$ for the TCGA lung cancer example.

\section*{Appendix C: Parameter Space for $\,\psi_I(p)$}

In this Appendix we describe the limits of parameter values for the density function $\,\psi_I(p\mid\bth)\,$ defined at~(\ref{psi.eq}) for small values of $\,I$.
Specifically, we must have $\,\psi_I(p)\,$ non-negative and monotone decreasing for all $\,0<p<1$.

For all values of $\,I\,$ we must have $\,\theta_I>0\,$ in order for $\,\psi_I(p)>0\,$ for values of $\,p\,$ close to zero.
We must have $\,\psi_I(1) = \theta_0\,$ non-negative so $\,\theta_0\geq 0$.

We also have $\,\psi'_I(1) = -\theta_1\,$ so for $\,\psi_I\,$ to be monotone decreasing, $\,\theta_1\geq 0\,$ for all values of $\,I$.
The condition that all $\,\theta_i\geq 0\,$ is sufficient (but not neccessary) for $\,\psi\,$ to be monotone decreasing because Descatres' rule states the derivative of $\,\psi(p)\,$ has no positive roots in $\,p$.

$\mbox{\boldmath $I=1\!:$}\,$ 
If $\,0\leq\theta_1\leq 1\,$ then $\,\psi_1(p\mid\theta_1)\,$ is a valid density and monotone decreasing.

$\mbox{\boldmath$I=2\!:$}\,$  
We must have $\,(\theta_0,\,\theta_1,\,\theta_2)\,$ all non-negative so
$$
    0 < \theta_2 \leq 1/2 {\rm \ \ and \ \ }
    0 \leq \theta_1 \leq 1-2\theta_2 \; . 
$$

\bigskip

For larger values of $\,I\!,$ define $\,x=-\log p\,$ and set $\,g(x) = \sum\theta_i x^i$.
It is sufficient for $\,g(x)\geq 0\,$ and $\,g'(x)\geq 0\,$ for all $\,x\geq 0\,$ to show $\,\psi\,$ is positive and monotone decreasing.
For $\,\theta_1\geq 0\,$ we have $\,g'(0)\geq 0\,$ and $\,g'(x)\geq 0\,$ for all $\,x\,$ sufficiently large because $\,\theta_I>0$.
To demonstrate $\,g'>0\,$ we need to show $\,g''(x)\,$ has no real, positive roots.

$\mbox{\boldmath$I=3\!:$}\,$   
We must have $\,\theta_3>0\,$ and $\,\theta_1\geq 0$.
The slope of $\,g(x)\,$ does not change sign provided its second derivative $\,g''=6\theta_3 x +2\theta_2\,$ is never negative for all $\,x  \geq 0$.
This shows $\,\theta_2>0$.
The restriction $\,0\leq\theta_0\leq 1\,$ gives
$$
     0 < \theta_3\leq 1/6; \quad\quad
     0\leq\theta_2\leq 1/2 - 3\theta_3 ; {\rm \ \ \ and \ \ \ }  
   0\leq\theta_1\leq 1 - 2\theta_2 - 6\theta_3 \; .
$$

$\mbox{\boldmath $I=4\!:$}\,$ We have $\,\theta_1\geq0\,$ and $\,\theta_4>0$. If the larger, real root of 
$\,g'' = 12\theta_4x^2 + 6\theta_3x + 2\theta_2\,$ is negative then
$$
  ( 36\theta_3^2 - 96\theta_2\theta_4 )^{1/2} < 6\theta_3
$$
showing $\,\theta_3>0$.  Squaring both sides of this inequality shows $\,\theta_2>0$.

If $\,g''\,$ has imaginary roots then 
$\, 36\theta_3^2 - 96\theta_2\theta_4 <0\,$ so $\,\theta_2>0\,$ and $\,g''\,$ is never negative.
With imaginary roots, if the minimum of $\,g''(x)\,$ occurs at $\,x>0\,$ then $\,\psi_4(p)\,$ will be decreasing but not concave.
The minimum of $\,g''(x)\,$ occurs at $\,x=-\theta_3/4\theta_4\,$ which is negative leading to $\,\theta_3>0.$

In either real or imagionary roots, for $\,I=4\,$ we have
\begin{eqnarray*}
  &0<\theta_4\leq1/24; \;\;\; 0\leq\theta_3\leq 1/6 - 4\theta_4;& \\ 
  & 0\leq\; \theta_2 \;\leq1/2 - 3\theta_3 - 12\theta_4;\;\;& \\
  & {\rm and \ \ \ } 0\leq\;\theta_1\;\leq 1-2\theta_2 - 6\theta_3 - 24\theta_4\; . &
\end{eqnarray*}

\end{document}